\documentclass[11pt,prd,superscriptaddress,preprintnumbers,amsmath,amssymb,nofootinbib]{revtex4-2}

\usepackage{epsfig}
\usepackage{bm}% bold math
\usepackage{dcolumn}% Align table columns on decimal point
\usepackage{textcomp}% Align table columns on decimal point
\usepackage{subfig}
\usepackage{booktabs}
\usepackage{color}
\usepackage{slashed}
\usepackage{soul}
\usepackage{braket}
\usepackage[]{hyperref}
\usepackage[table]{xcolor}
  \hypersetup{
  unicode=false,          % non-Latin characters in Acrobat?s bookmarks
  pdftoolbar=true,        % show Acrobat?s toolbar?
  pdfmenubar=true,        % show Acrobat?s menu?
  pdffitwindow=true,     % window fit to page when opened
  pdfstartview={FitH},    % fits the width of the page to the window
  pdfnewwindow=true,      % links in new window
  pdfcreator={RevTeX},
  colorlinks=true,       % false: boxed links; true: colored links
  linkcolor=red,          % color of internal links
  citecolor=blue,        % color of links to bibliography
  urlcolor=blue,           % color of external links
  }
\usepackage{hypcap}

\newcommand{\real}{\mathrm{Re}\,}

\large

\begin{document} 

\title{Impact of $b \to c$ measurements on $\Lambda_b \to p \tau \bar{\nu}$ decay in $U_1$ leptoquark model}

\author{Shabana Khan}
\email{shabana26k@gmail.com}
\affiliation{Department of Physics, University of Rajasthan, Jaipur 302004, India}

\author{Neetu Raj Singh Chundawat}
\email{chundawat.1@iitj.ac.in}
\affiliation{Indian Institute of Technology Jodhpur, Jodhpur 342037, India}

\author{Dinesh Kumar}
\email{dineshsuman09@gmail.com}
\affiliation{Department of Physics, University of Rajasthan, Jaipur 302004, India}

\begin{abstract}
The measurements of several lepton flavor universality (LFU) violating observables in the decays induced by the quark level transition $b \to c \tau \bar{\nu}$ provide an inkling of plausible physics beyond the standard model of electroweak interactions. Such  new physics would also impact other sectors. In this work, we estimate the leverage of new physics in $b \to c \tau \bar{\nu}$ on  $\Lambda_b \to p \tau \bar{\nu}$ decay in the context of $U_1$ leptoquark model. In this model, the new physics couplings in $b \to u \tau \bar{\nu}$ transition can be written  in terms of $b \to c \tau \bar{\nu}$ couplings and hence the extent of allowed new physics in $\Lambda_b \to p \tau \bar{\nu}$  would be determined by  $b \to c \tau \bar{\nu}$ transition. Using the new physics parameter space obtained by  performing a fit to all $b \to c \tau \bar{\nu}$ data, we obtain predictions of several $\Lambda_b \to p \tau \bar{\nu}$  observables. We find that the current $b \to c \tau \bar{\nu}$ data allows two times of magnitude enhancement in the branching ratio as well as in the LFU ratio. The other observables such as convexity parameter, lepton forward-backward asymmetry, longitudinal polarization of final state baryon and tau lepton are consistent with the SM value.
\end{abstract}
 
\maketitle 

\newpage
 
\section{Introduction}
The standard model (SM) of electroweak interactions can be considered as a baronial theory of fundamental interactions of nature. Ever since the discovery of weak neutral currents in 1973 in a neutrino scattering experiment in the Gargamelle bubble chamber at CERN, SM has been substantiated through a fecundity of experimental observations. The discovery of the Higgs Boson marks the culmination of the particle spectrum of the SM. Though SM may flare out to be an irrefutable theory, there are several other observations which propel us to clamor for physics beyond the SM. These include disappearance of anti-matter, existence of dark matter and dark energy. Further, gravity is excluded from the SM. Therefore the nonpareil theory of fundamental interactions of nature is still far away from the bay.

The evidence of physics beyond the SM has already started burgeoning at several fronts. These include observables related to the decays of $B$ mesons. These anomalous discrepancies can be classified into two categories: decays induced by the charged current transition $b \to c \ell \nu$ ($\ell=e,\,\mu,\, \tau$) and neutral current transition $b \to s \ell \ell$ ($\ell=e,\,\mu$). In this work, we rivet on decays induced by the $b \to c \ell \nu$ transition which occurs at the tree level in the SM. A series of measurements by the Belle, BaBar and LHCb collaborations over the last decade have provided several enthralling hints of new physics in this sector.

The BaBar~\cite{Lees:2012xj,Lees:2013uzd}, Belle~\cite{Huschle:2015rga,Sato:2016svk,Hirose:2016wfn} and LHCb~\cite{Aaij:2015yra,lhcb-2023,hflav-2023} collaborations measured the following flavor ratios
\begin{equation}
R_{D^{(*)}}\equiv \frac{{\Gamma}(B \to D^{(*)}\, \tau\, \bar \nu)}{{\Gamma}(B \to D^{(*)}\, (e,\, \mu) \,\bar \nu)}.
\end{equation}
The average values of these measurements differ from their respective SM predictions at the level of 3.2$\sigma$~\cite{hflav-2023}.
These deviations are inklings of lepton flavor universality violation. All of these experiments were based on methodologies where the $\tau$ lepton was identified through kinematical information rather than reconstruction. The reconstruction technique was emplaced by the LHCb collaboration using the $3\pi$ decay mode of the $\tau$ lepton ~\cite{Aaij:2017uff}. This resulted in a distinct measurement of  $R_{D^*}$. Including this measurement, the incongruence of the
$R_D$-$R_{D^*}$ data with SM predictions  escalated to $4.1\sigma$~\cite{hflav-2017}. 
In 2019, Belle collaboration announced a new measurement of $R_D$ and $R_{D^*}$~\cite{Abdesselam:2019dgh}, which is consistent with the SM prediction. Very recently, on March 21, 2023, the LHCb collaboration updated the value of $R_{D^{*}}$. By including these measurements, the discrepancy with SM reduced from $4.1\sigma$ to $3.2\sigma$. 

Apart from  $R_{D^{(*)}}$, the  LHCb collaboration measured the following ratio  in $B_c\rightarrow J/\psi \, \ell \, \bar{\nu}$ decay modes
\begin{equation}
R_{J/\psi} = \frac{\Gamma(B_c\rightarrow J/\psi \, \tau \, \bar{\nu})}{\Gamma(B_c\rightarrow J/\psi \, \mu\, \bar{\nu})}\,.
\end{equation} 
They found that the measured value is $R_{J/\Psi} = 0.71 \pm 0.17 \pm 0.18$~\cite{Aaij:2017tyk}. This decay is also generated by the same quark level transition which induces $R_{D^{(*)}}$. The measured value is $1.8\sigma$ higher than the latest SM prediction of $0.2582(38)$~\cite{Harrison:2020nrv}. These dissension with  the SM can be imputed to new physics in $\tau$, $\mu$ or $e$ sectors. However, in \cite{Alok:2017qsi} it was shown that new physics only in $\mu$ or $e$ sectors cannot accommodate these measurements. This is mainly due to measurements of the ratios  $
R^{\mu/e}_D = \Gamma(B\rightarrow D\,\mu\,\nu)/\Gamma(B\rightarrow D\,e\,\nu)= 0.995\pm 0.022\, {\rm(stat.)}\pm 0.039\, {\rm (syst.)}$ and 
$
R^{e/\mu}_{D^*} = \Gamma(B\rightarrow D^* \,e \, \nu)/\Gamma(B\rightarrow D^* \,\mu \, \nu) = 1.04\pm 0.05\, {\rm(stat.)}\, \pm 0.01  {\rm (syst.)}
$ \cite{Glattauer:2015teq, Abdesselam:2017kjf}.
The measured values of these ratios are in agreement with their SM predictions. Hence new physics only in $b\rightarrow c\,\mu\,\bar{\nu}$ or $b\rightarrow c\, e \,\bar{\nu}$ will blight this agreement.  Therefore new physics in $b \to c \tau  \nu$ is imperative to accommodate the current measurements of flavor ratios in these sectors \footnote{In \cite{Carvunis:2021dss}, it was shown that new physics only in muons can accommodate the entire  $b\rightarrow c\, l \,\bar{\nu}$ data using a different set of combinations of new physics operators. }.  

In May 2022, the LHCb collaboration reported the first observation of the semileptonic $b$-baryon decay $\Lambda_b \to \Lambda_c^+ \tau^- \bar{\nu_{\tau}}$ with a significance of 6.1$\sigma$ \cite{LHCb:2022piu}. This was obtained by collecting a data sample corresponding to 3 $\rm fb^{-1}$ of integrated luminosity at centre-of-mass energies of 7 and 8 TeV. The LFU ratio $R(\Lambda_c)$ was measured to be \cite{LHCb:2022piu}
\begin{eqnarray}
R(\Lambda_c) &=&\frac{Br(\Lambda_b \to \Lambda_c^+ \tau^- \bar{\nu_{\tau}})}{Br(\Lambda_b \to \Lambda_c^+ \mu^- \bar{\nu_{\mu}})}\nonumber \\
&=&  0.242 \pm 0.026 \,(\rm stat.) \pm 0.040 \,(\rm syst.) \pm 0.059\,.
\end{eqnarray}
Here  the last error is due to the external branching fraction uncertainty
from the channel $\Lambda_b \to \Lambda_c^+ \mu^- \bar{\nu_{\mu}}$. The measured value is consistent with the SM prediction  of $0.324 \pm  0.004$ \cite{Bernlochner:2018bfn}. 

Barring these LFU observables, we also have measurements of few angular observables.  The Belle collaboration has measured the $\tau$ polarization, $P^{D^*}_{\tau}$, in $B \to D^* \tau \bar{\nu}$ decay. The measured value \cite{Hirose:2016wfn}
\begin{equation}
P^{D^*}_{\tau} =  - 0.38 \pm 0.51\, (\rm stat.) ^{+0.21}_{-0.16}\, (\rm syst.),
\end{equation}
is consistent with its SM prediction of $-0.497\pm0.013$ \cite{Tanaka:2012nw}. In 2018,
Belle collaboration reported the measurement of $D^*$ longitudinal polarization fraction $F^{D^*}_L$ in the decay $B \to D^* \tau \bar{\nu}$.  The measured value  \cite{Abdesselam:2019wbt}
\begin{equation}
F^{D^*}_{L} = 0.60 \pm 0.08\, (\rm stat.)\pm 0.04\, (\rm syst.)\,
\end{equation} is $1.6\sigma$ higher than the SM prediction of $0.46\pm 0.04$~\cite{Alok:2016qyh}.

The possible new physics effects in  $b\rightarrow c\tau\bar{\nu}$  decay can be analyzed in a model independent way using the language of effective field theory. There are many such analyses, see for e.g, \cite{Freytsis:2015qca,Jung:2018lfu,Bhattacharya:2018kig,Hu:2018veh,Alok:2019uqc,Asadi:2019xrc,Murgui:2019czp,Bardhan:2019ljo,Blanke:2019qrx,Shi:2019gxi,Becirevic:2019tpx,Sahoo:2019hbu,Cheung:2020sbq,Cardozo:2020uol}. These analyses identified Lorentz structure of possible new physics. However, there are no unique solutions. Depending upon the adopted methodology and assumptions, there are multiple new physics operators with specific values of corresponding WCs which can provide a good fit to data. A unique determination of the new Lorentz structure of new physics would require measurements of additional observables in $b\rightarrow c\tau\bar{\nu}$ sector \cite{Alok:2018uft}.

The allowed model independent solutions can be realized in specific new physics models. There are a good number of such models. In context of some of these models it would be interesting to see whether some correlations exist between the observables in $b \to c$ sector and other sectors. In other words, what implications measurement in $b \to c$ sector have on other sectors. In this work we explore such correlations in $b \to u$ sector in the context of $U_1$ leptoquark (LQ) model. The $U_1$ leptoquark is extensively discussed in the literature in the context of B-anomalies\cite{Feruglio:2017rjo,Calibbi:2017qbu,Blanke:2018sro,Crivellin:2018yvo,Kumbhakar:2020okw,Cornella:2021sby,Bernigaud:2021fwn}. In particular, we study imprints of $b \to c$ measurements on several observables in $\Lambda_b \to p \tau \bar{\nu}$ decay mode. The baryonic decay mode $\Lambda_b \to p \tau \bar{\nu}$ is studied in the literaure \cite{Dutta:2015ueb,Ray:2018hrx}. 

The quark level transition  $b\rightarrow u\tau\bar{\nu}$ induces decays such as $B^+ \to \tau \bar{\nu}$, $B \to \pi  \tau \bar{\nu}$, $B \to \rho  \tau \bar{\nu}$, $B \to \omega  \tau \bar{\nu}$ and  $\Lambda_b \to p \tau \bar{\nu}$. Out of these decays, currently, the only observed decay channel is the purely leptonic decay $B^+ \to \tau \bar{\nu}$ \cite{pdg}. The measured value of its branching ratio is $(1.09 \pm 0.24) \times 10^{-4}$ which is consistent with the SM value $(9.89 \pm 0.13) \times 10^{-5}$\cite{Cardozo:2020uol}. Further, the Belle collaboration provides an upper bound on the branching ratio of the semileptonic decay $B \to \pi  \tau \bar{\nu}$. At 90\% C.L., the branching ratio of  $B \to \pi  \tau \bar{\nu}$ can be as high as $2.5 \times 10^{-4}$ \cite{Belle:2015qal}. Thus due to lack of enough  measurements, any model independent analysis would allow a large new physics effects in some of the observables in  $b\rightarrow u\tau\bar{\nu}$ transition. In other words, given the current experimental situation in $b\rightarrow u\tau\bar{\nu}$ sector, a model dependent framework will engender more meaningful analysis as compared to the model independent analysis in the sense that it would allow for additional constraints coming from other sectors. In the context of $U_1$ leptoquark model considered in this work, we show that the necessary couplings in $b\rightarrow u\tau\bar{\nu}$ decay are all related to the couplings  in $b\rightarrow c\tau\bar{\nu}$ sector. Given the fact that we have relatively accurate measurements of number of observables in this sector, it would be interesting to see the extent up to which the new physics effects are allowed in $b\rightarrow u\tau\bar{\nu}$ sector. In particular, we study the impact of $b\rightarrow c\tau\bar{\nu}$ measurements on  several observables in $\Lambda_b \to p \tau \bar{\nu}$ decay mode.

Plan of work is as follows. In Sec.\ref{tf}, we provide theoretical framework of this work. Starting with the effective Hamiltonian, we provide all necessary theoretical expressions in this section. This includes various observables in  $b \to c\, \tau \, \bar{\nu}$  sector and  $\Lambda_b \to p \tau \bar{\nu}$ decay. In the next section, we  first provide constraints on  $b \to c\, \tau \, \bar{\nu}$ couplings by performing a fit. Using the allowed parameter space of these couplings, we obtain predictions of several observables in $\Lambda_b \to p \tau \bar{\nu}$ decay. The conclusions are discussed in Sec.\ref{concl}.

%%%%%%%%%%%%%%%%%%%%%%%%%%%%%%%%%%%
\section{Theoretical Framework}
\label{tf}
%%%%%%%%%%%%%%%%%%%%%%%%%%%%%%%%%%%

%%%%%%%%%%%%%%%%%%%%%%%%%%%%%%%%%%%
\subsection{Effective Hamiltonian}
\label{eff}
%%%%%%%%%%%%%%%%%%%%%%%%%%%%%%%%%%%

Within the SM, the effective Hamiltonian for the quark level transition $b \to q\, \tau \, \bar{\nu}$ with $q = u,c$ is given by 
\begin{equation}
H_{eff}^{\rm SM} = \frac{4 G_F}{\sqrt{2}} V_{qb} \,O_{V_L}\, +\,\it h.c. ,
\label{effH}
\end{equation}
where ${O}_{V_L} =(\bar{q} \gamma_\mu P_L b)\,(\bar{\tau} \gamma^\mu P_L \nu)$.  In the presence of new physics, the effective Hamiltonian takes the form
\begin{equation}
H_{eff} = \frac{4 G_F}{\sqrt{2}} V_{qb} [ (1+C_{V_L})  O_{V_L}     +  C_{V_R}  O_{V_R}   +  C_{S_L}  O_{S_L}  +  C_{S_R}  O_{S_R} + C_T  O_T     ]\, +\,\it h.c. ,
\end{equation}
where
\begin{eqnarray}
{O}_{V_R}   &=& (\bar{q} \gamma_\mu P_R b)\,(\bar{\tau} \gamma^\mu P_L \nu)\,,\\
  {O}_{S_R}&=& (\bar{q} P_R b)\,(\bar{\tau} P_L \nu)\,,\\
  {O}_{S_L}&=& (\bar{q} P_L b)\,(\bar{\tau} P_L \nu)\,,\\
  {O}_T&=& (\bar{q}\sigma^{\mu\nu}P_L b)\,(\bar{\tau}\sigma_{\mu\nu}P_L \nu)\,.
\end{eqnarray}

The interactions between the vector singlet $U_1$ LQ and the SM quarks and leptons can be written as \cite{Dorsner:2016wpm,Bhaskar:2021pml}
\begin{equation}
H^{U_1} = h^L_{ij}\bar{Q}^i \gamma_{\mu}U_1^{\mu}P_L L^j + h^R_{ij}\bar{d}^i \gamma_{\mu}U_1^{\mu}P_R l^j_R + {\it h.c.,}
\end{equation}
where $Q_i$ and $L_j$ are the SM left-handed quark and lepton doublets and $d_R^i$ and $l^j_R$  are right handed quarks and leptons.
Here $h^L_{ij}$ and $h^R_{ij}$ are the 3$\times$ 3 matrices in the flavor space. This LQ contributes to $b \to c \tau \bar{\nu}$ at the tree level. As we only require $\bar{c}\,\nu \,U_1$ and $\bar{b}\, \tau\, U_1$ couplings to be non-zero, we have 
\begin{equation}
 h^L=\begin{pmatrix}
0 & 0 & 0\\
0 & 0 & h^L_{23}\\
0 & 0 & h^L_{33}
\end{pmatrix}\,,
\quad \quad
 h^R=\begin{pmatrix}
0 & 0 & 0\\
0 & 0 & 0\\
0 & 0 & h^R_{33}
\end{pmatrix}
\end{equation}
Assuming mixing in the up-type quark sector, the interaction Hamiltonian in the physical quarks can be written by rotating them with the CKM matrix and is given as 
\begin{eqnarray}
H_{ eff} &=& \Bigg[\left(V_{us}h_L^{23}+V_{ub}h_{33}^L \right)\bar{u_L}\gamma_{\mu}\nu_L + \left(V_{cb}h_L^{33}+V_{cs}h_{23}^L \right)\bar{c_L}\gamma_{\mu}\nu_L \nonumber\\
&& + h_L^{23} \bar{s}_L\gamma_{\mu}\tau_L + h_L^{33} \bar{b}_L\gamma_{\mu}\tau_L  + h_R^{33} \bar{b}_R\gamma_{\mu}\tau_R \Bigg]U_1^{\mu}\, + {\it h.c.}
\end{eqnarray}
It is ostensible from the above Lagrangian that only $O_{V_L}$ and $O_{S_R}$ contribute to 
$b \to c \tau \bar{\nu}$ and $b  \to u \tau \bar{\nu}$ processes. Also, the same couplings appear in both decay modes.  The relevant WCs for $b \to c \tau \bar{\nu}$ decay can be written as
\begin{eqnarray}
C_{V_L}^{b \to c} &=& \frac{1}{2\sqrt{2}G_F V_{cb}} \frac{\left(V_{cb}h^L_{33}+V_{cs}h^L_{23}\right)h_{33}^L}{M^2_{U_1}}\,,\\
C_{S_R}^{b \to c} &=& -\frac{1}{\sqrt{2}G_F V_{cb}} \frac{\left(V_{cb}h^L_{33}+V_{cs}h^L_{23}\right)h_{33}^R}{M^2_{U_1}}\,.
\end{eqnarray}

The WCs for  $b \to u \tau \bar{\nu}$ decay are
\begin{eqnarray}
C_{V_L}^{b \to u} &=& \frac{1}{2\sqrt{2}G_F V_{ub}} \frac{\left(V_{ub}h^L_{33}+V_{us}h^L_{23}\right)h_{33}^L}{M^2_{U_1}}\,,\\
C_{S_R}^{b \to u} &=& -\frac{1}{\sqrt{2}G_F V_{ub}} \frac{\left(V_{ub}h^L_{33}+V_{us}h^L_{23}\right)h_{33}^R}{M^2_{U_1}}\,.
\end{eqnarray}
Thus we see that the $b \to c$ couplings can determine the new physics contributions to $b \to u$. Therefore we need to analyze observables in the $b \to c\, \tau \, \bar{\nu}$ sector. The new physics scalar effective Wilson coefficients are affected by the QCD running from the TeV scale down to the $m_b$ scale and it is considered in our analysis. It should be noted that the considered $U_1$ leptoquark model  is non-renormalizable and hence it requires a UV completion.

In the next section we provide theoretical expressions for $b \to c\, \tau \, \bar{\nu}$ observables used in our analysis to constrain the new physics parameter space. 

%%%%%%%%%%%%%%%%%%%%%%%%%%%%%%%%%%%
\subsection{Observables in $b \to c\, \tau \, \bar{\nu}$ sector}
\label{b2c}
%%%%%%%%%%%%%%%%%%%%%%%%%%%%%%%%%%%

We consider following observables in our analysis:
\begin{itemize}
\item the flavor ratios $R_D$, $R_{D^*}$, $R_{J/\Psi}$  and $R(\Lambda_c)$,

\item tau polarization in $B \to D^* \tau \bar{\nu}$  decays,

\item $D^*$ longitudinal polarization fraction in $B \to D^* \tau \bar{\nu}$ decay,

\item branching ratio of $B_c\rightarrow \tau\bar{\nu}$.
\end{itemize}

The theoretical expressions for $R_D$, $R_{D^*}$ and $   R_{\Lambda_c}$ in terms of WCs  are given as \cite{Blanke:2018yud}
\begin{eqnarray}
    R^{th}_D   &\simeq & R_D^{\rm SM}  \big\{\vert 1+C_{V_L}\vert^2+1.54 \,\real[(1+C_{V_L}) C_{S_R}] +1.09 \vert C_{S_R}\vert^2\big\}\,, \label{eq:rd}\\
    R^{th}_{D^*} &\simeq& R_{D^*}^{\rm SM}    \big\{\vert 1+C_{V_L}\vert^2 + 0.13\,\real[(1+C_{V_L})C_{S_R}] +0.05\vert C_{S_R}\vert^2\big\} \,,\label{eq:rds}\\
   R^{th}_{\Lambda_c}   &\simeq&  R_{\Lambda_c}^{\rm SM}  \big\{\vert 1 + C_{V_L}\vert^2  + 0.50 \,\real[(1 + C_{V_L}) C_{S_R} ] +0.33  \vert  C_{S_R} \vert^2 \big\}.
 \label{eq:rlc}
\end{eqnarray}

We used the form factors computed in the full $q^2$ range using Lattice QCD~\cite{Harrison:2020gvo} and obatined the theoretical expression for $R_{J/\Psi}$ in terms of the NP WCs which is given by 
\begin{equation}
R^{th}_{J/\Psi}   \simeq   0.2581\vert 1+C_{V_L}\vert^2+0.027 \,\real[(1+C_{V_L}) C_{S_R}]+0.01 \vert C_{S_R}\vert^2 \,.
\end{equation}

Tau polarization in $B \to D^* \tau \bar{\nu}$  decay, $P_\tau^{D^*}$,  in the $U_1$ LQ model is given as \cite{Blanke:2018yud} 
\begin{equation}
    P_{\tau}^{D^*\, th}  \simeq \left(\frac{R^{th}_{D^*}}{R_{D^*}^{\rm SM} }\right)^{-1}\Big\{-0.49 \vert1 + C_{V_L}\vert^2 + 0.05 \vert C_{S_R}\vert^2  +0.13 \,\real[(1 +C_{V_L}) C_{S_R}] \Big\}\,.
\end{equation}

The expression for $D^*$ longitudinal polarization fraction,  $f_L^{D^*}$ ,  in $B \to D^* \tau \bar{\nu}$ decay is \cite{Blanke:2018yud} 
\begin{equation}
f_L^{D^*\, th}  \simeq  \left(\frac{R^{th}_{D^*}}{R_{D^*}^{\rm SM} }\right)^{-1}\Big\{0.46 \vert1 + C_{V_L}\vert^2 + 0.05 \vert  C_{S_R}\vert^2   + 0.13 \,\real[(1 + C_{V_L}) C_{S_R}]\Big\}. 
\end{equation}

We also consider the constraints coming  from the purely leptonic decay $B_c\rightarrow \tau\,\bar{\nu}$. The branching ratio of $B_c$ is used to check the consistency of the fit results.
This decay mode is not affected by the helicity suppression provided the transition is induced through the  pseudo-scalar operators. The branching ratio of $B_c\rightarrow \tau\bar{\nu}$ in the $U_1$ LQ model can be written as
\begin{eqnarray}
{\cal B}(B_c\rightarrow \tau\bar{\nu})& \simeq & 0.02\bigg(\frac{f_{B_c}}{0.43\,\text{GeV}}\bigg)^2 \Big\vert 1+C_{V_L} + 4.3\,C_{S_R}\Big\vert ^2.
\end{eqnarray}

%%%%%%%%%%%%%%%%%%%%%%%%%%%%%%%%%%%
\subsection{Observables in $\Lambda_b \to p \tau \bar{\nu}$  decay mode}
\label{b2u}
%%%%%%%%%%%%%%%%%%%%%%%%%%%%%%%%%%%
In this section, we provide theoretical expressions for various $\Lambda_b \to p \tau \bar{\nu}$ observables used in our analysis. These observables can be defined with the help of angular differential decay distribution of this mode.
The two-fold angular differential distribution for $\Lambda_b \to p l\bar{\nu}$ can be written in terms of $q^2$ and $\cos\theta_l$ where $q^2$ is the momentum transfer squared and $\theta_l$ is the angle between the daughter baryon and the lepton in the di-lepton rest frame. The two-fold angular differential distribution can be written as 
\begin{equation}
\frac{d^2\Gamma(\Lambda_b \to p l\bar{\nu})}{dq^2\, d\cos\theta_l} = N \Big(1-\frac{m_l^2}{q^2}\Big)^2\Big[A + \frac{m_l^2}{q^2}B + 2\,C + \frac{4m_l}{\sqrt{q^2}}D\Big]\,,
\label{twofold}
\end{equation}
where 
\begin{eqnarray}
A &=& 2\sin^2\theta_l\Big(H^2_{\small{\frac{1}{2},0}} + H^2_{\small{-\frac{1}{2},0}}\Big) + \Big(1-\cos\theta_l\Big)^2 H^2_{\small{\frac{1}{2},1}}
+ \Big(1+\cos\theta_l\Big)^2H^2_{\small{-\frac{1}{2},-1}},\\
B &=& 2\cos^2\theta_l\Big(H^2_{\small{\frac{1}{2},0}} + H^2_{\small{-\frac{1}{2},0}}\Big) + \sin^2\theta_l \Big(H^2_{\small{\frac{1}{2},1}} + H^2_{\small{-\frac{1}{2},-1}}\Big)+ 2\Big(H^2_{\small{\frac{1}{2},t}} + H^2_{\small{-\frac{1}{2},t}}\Big)  \nonumber \\
&&- 4\cos\theta_l\Big( H_{\small{\frac{1}{2},t}} H_{\small{\frac{1}{2},0}} +  H_{\small{-\frac{1}{2},t}} H_{\small{-\frac{1}{2},0}}\Big) \\
C &=& \Big(H^{SP}_{\small{\frac{1}{2},0}}\Big)^2 + \Big(H^{SP}_{\small{-\frac{1}{2},0}}\Big)^2,\\
D &=& -\cos\theta_l\Big(H_{\small{\frac{1}{2},0}} H^{SP}_{\small{\frac{1}{2},0}} + H_{\small{-\frac{1}{2},0}} H^{SP}_{\small{-\frac{1}{2},0}}\Big) + \Big(H_{\small{\frac{1}{2},t}} H^{SP}_{\small{\frac{1}{2},0}} + H_{\small{-\frac{1}{2},t}} H^{SP}_{\small{-\frac{1}{2},0}}\Big)\,.
\end{eqnarray}

The differential decay rate for $\Lambda_b \to p l\bar{\nu}$ can be obtained after integrating out equation \ref{twofold} over the $\cos\theta_l$ variable \cite{Shivashankara:2015cta}
\begin{equation}
\frac{d\Gamma(\Lambda_b \to p l\bar{\nu})}{dq^2} = \frac{8N}{3}\Big(1-\frac{m_l^2}{q^2}\Big)^2\Big[E + \frac{m_l^2}{2q^2}F + \frac{3}{2}G + \frac{3m_l}{\sqrt{q^2}}H\Big]\,.
\end{equation} 
Here $N = \frac{G_F^2|V_{ub}|^2 q^2|\vec{p_p}|}{512\pi^3 m_{\Lambda_b}^2}, |\vec{p_p}| = \sqrt{\lambda(m^2_{\Lambda_b},m_p^2,q^2)}/(2m_{\Lambda_b})$ with $\lambda(a,b,c) = a^2 + b^2 + c^2 - 2(ab+bc+ca)$. Further, 
\begin{eqnarray}
E &=& H^2_{\small{\frac{1}{2}0}} + H^2_{\small{-\frac{1}{2}0}} + H^2_{\small{\frac{1}{2}1}} + H^2_{\small{-\frac{1}{2}-1}},\\
F &=& H^2_{\small{\frac{1}{2}0}} + H^2_{\small{-\frac{1}{2}0}} + H^2_{\small{\frac{1}{2}1}} + H^2_{\small{-\frac{1}{2}-1}} + 3(H^2_{\small{\frac{1}{2}t}} + H^2_{\small{-\frac{1}{2}t}}),\\
G &=& (H^{SP}_{\small{\frac{1}{2}0}})^2 + (H^{SP}_{\small{-\frac{1}{2}0}})^2,\\
H &=& H_{\small{\frac{1}{2}t}} H^{SP}_{\small{\frac{1}{2}0}} + H_{\small{-\frac{1}{2}t}} H^{SP}_{\small{-\frac{1}{2}0}}\,.
\end{eqnarray}
The differential branching fraction can then be written as 
\begin{equation}
\frac{d\mathcal{B}(\Lambda_b \to p l \bar{\nu})}{dq^2} = \tau_{\Lambda_b} \frac{d\Gamma}{dq^2}\,.
\label{dbr}
\end{equation}
One can also define the following LFU ratios of the differential branching fractions as
\begin{equation}
R_p(q^2) = \frac{d\Gamma(\Lambda_b \to p \tau\bar{\nu})/dq^2}{d\Gamma(\Lambda_b \to p \mu\bar{\nu})/dq^2},\,\,\,\,\,
\label{rlfu}
\end{equation}
The lepton forward-backward asymmetry is defined as
\begin{equation}
A_{FB} = \frac{\int_{0}^{1} (d^2\Gamma/dq^2\,d\cos\theta)d\cos\theta - \int_{-1}^{0} (d^2\Gamma/dq^2\,d\cos\theta)d\cos\theta}{\int_{0}^{1} (d^2\Gamma/dq^2\,d\cos\theta)d\cos\theta + \int_{-1}^{0} (d^2\Gamma/dq^2\,d\cos\theta)d\cos\theta}\,.
\label{afb}
\end{equation}
Moreover, the longitudinal polarization of final state baryon and $\tau$ lepton is given by
\begin{eqnarray}
P^L_{p} &=& \frac{d\Gamma^{\lambda_p = 1/2}/dq^2 - d\Gamma^{\lambda_p = -1/2}/dq^2}{d\Gamma^{\lambda_p = 1/2}/dq^2 + d\Gamma^{\lambda_p = -1/2}/dq^2}\\
P^L_{\tau} &=& \frac{d\Gamma^{\lambda_{\tau} = 1/2}/dq^2 - d\Gamma^{\lambda_{\tau} = -1/2}/dq^2}{d\Gamma^{\lambda_{\tau} = 1/2}/dq^2 + d\Gamma^{\lambda_{\tau} = -1/2}/dq^2}
\label{pol}
\end{eqnarray}
The convexity parameter, which is the measure of curvature of the $\cos\theta$ distribution, is defined as
\begin{equation}
C_F^l(q^2) = \frac{1}{\int d\cos\theta\, W(\theta)}\frac{d^2 W(\theta)}{d(\cos\theta)^2}
\label{conv}
\end{equation}
with 
$$W(\theta) = \frac{3}{8}\Big[A + \frac{m_l^2}{q^2} B +2\,C + \frac{4m_l}{\sqrt{q^2}}D\Big].$$
The helicity amplitudes defined in terms of the form factors are given in Appendix \ref{appen}.
%%%%%%%%%%%%%%%%%%%%%%%%%%%%%%%%%%%
\section{Results and Discussions}
\label{res}

%%%%%%%%%%%%%%%%%%%%%%%%%%%%%%%%%%%
\subsection{Fit results}
\label{fit}
%%%%%%%%%%%%%%%%%%%%%%%%%%%%%%%%%%%

From Sec \ref{eff}, it is apparent that in the context of $U_1$ LQ model, the WCs in $b \to u$ transition can be written in terms of $b \to c$ couplings. Therefore the observables in $b \to u $ sector are expected to have strong correlations with  $b \to c$ observables. In other words, the extent up to which the new physics effects can be generated in $b \to u$ observables would be determined by the allowed parameter space of couplings by the current $b \to c$ data. Given the fact that we have relatively large number of measured observables in this sector and moreover, some of them are accurately measured and predicted fairly well within the SM, it would be interesting to see possible deviation in $\Lambda_b \to p \tau \bar{\nu}$ observables  allowed by the  $b \to c$ data.

{\rowcolors{2}{black!50!white!50}{black!50!white!40}
\begin{table}
\centering
\begin{tabular}{ |c|c| }
\hline
Observable& Experimental Values \\
\hline
$ R_D$     & $ 0.356\pm 0.029$   \cite{hflav-2023} \\
$R_{D^*} $  & $0.284\pm 0.013$ ~\cite{hflav-2023}\\
$R_{J/\Psi} $  & $ 0.71\pm 0.17\pm 0.18$ ~\cite{Aaij:2017tyk}\\
$R_{\Lambda_c}  $  &   $ 0.242 \pm 0.026 \,(\rm stat.) \pm 0.040 \,(\rm syst.) \pm 0.059 $  \cite{LHCb:2022piu} \\
$ P_{\tau}^{D^*}$  & $-0.38\pm 0.51^{+0.21}_{-0.16} $  \cite{Hirose:2016wfn} \\
$ f_L^{D^*}$  &  $0.60\pm 0.08\,(\rm stat.)\pm 0.04\,(\rm syst.) $  \cite{Adamczyk:2019wyt,Abdesselam:2019wbt} \\
\hline
\end{tabular}
\caption{Experimental values of observables used in the fit. The third error in $R_{\Lambda_c}  $ is due to the external branching fractions measurements.}
\label{fit-obs}
\end{table}
% \frac{(R_{J/\Psi}^{th}(C_i)-R_{J/\Psi}^{exp})^2}{\sigma^2_{R_{J/\Psi}}+

The theoretical expressions of  observables $R_D$, $R_{D^*}$,  $R_{\Lambda_c}$ , $P_\tau^{D^*}$ and $f_L^{D^*}$ as functions of the relevant WCs are given in Sec. \ref{b2c}. By fitting these expressions to the measured values of the observables, we obtain the values of WCs which
are consistent with the data.
The corresponding $\chi^2$ is defined as
\begin{small}
\begin{eqnarray}
\chi^2(C^{\rm{eff}}_i)&=&\sum_{m,n= R_D, R_{D^*}}\left(O^{th}(C_i)-O^{exp}\right)_{m}\left(V\right)^{-1}_{mn}\left(O^{th}(C_i)-O^{exp}\right)_{n}\nonumber\\
& &+ \frac{(R_{J/\Psi}^{th}(C_i)-R_{J/\Psi}^{exp})^2}{\sigma^2_{R_{J/\Psi}}}+ \frac{(R_{\Lambda_c}^{th}(C_i)-R_{\Lambda_c}^{exp})^2}{\sigma^2_{R_{\Lambda_c}}} \nonumber \\
& &+ \frac{(P_{\tau}^{D^*\, th}(C_i)-P_{\tau}^{D^*\,exp})^2}{\sigma^2_{P_{\tau}}}  + \frac{(f_L^{D^*\, th}(C_i)-f_L^{D^*\,exp})^2}{\sigma^2_{f_L}}.
\label{chi2}
\end{eqnarray}
\end{small}

where $V = V^{exp} + V^{SM}$. Here $O^{th}(C^{\rm{eff}}_i)$ are the theoretical predictions for $R_D$ and $R_{D^*}$ whereas $R_{J/\Psi}^{th}$, $R_{\Lambda_c}^{th}$ , $P_{\tau}^{D^*\, th}$ and $f_L^{D^*\, th}$ are theoretical expressions for $R_{J/\Psi}$,  $R_{\Lambda_c}$, $P_{\tau}^{D^*}$ and $f_L^{D^*}$, respectively. These expressions depend upon the new physics WCs $C_{V_L}$ and $C_{S_R}$ which in turn are functions of $h_{23}^L$, $h_{33}^L$ and $h_{33}^R$ couplings. $O^{exp}$ are the corresponding experimental measurements. $V^{exp}$ and $V^{SM}$ are the experimental and SM covariance matrices in the $R_D$, $R_{D^*}$ space, respectively. The matrix $V^{exp}$ includes the correlation in the combined experimental determination of $R_D$ and $R_{D^*}$. In eq.~(\ref{chi2}), $\sigma_{R_{J/\Psi}}$, $\sigma_{R_{\Lambda_c}}$, $\sigma_{P_{\tau}}$ and , $\sigma_{f_L}$ are the uncertainties in the measurements and theory (added in quadrature) of $R_{J/\Psi}$, $R_{\Lambda_c}$ , $P_\tau^{D^*}$ and $f_L^{D^*}$, respectively. The measured values are given in Table \ref{fit-obs}.

We now consider three different scenarios by fixing one of the couplings and varying the remaining two. These scenarios are as follows:
\begin{itemize}
\item $S1$: $h_L^{33} = 0.5, M_{U_1} = 1.5 $ TeV and varying $h_L^{23} \,\,\&\,\, h_R^{33}$.
\item $S2$: $h_L^{23} = 0.5, M_{U_1} = 1.5 $ TeV and varying $h_L^{33} \,\,\&\,\, h_R^{33}$.
\item $S3$ : $h_R^{33} = 0.5, M_{U_1} = 1.5 $ TeV and varying $h_L^{33}\,\, \&\,\, h_L^{23}$.
\end{itemize}
The best fit values for these three scenarios are shown in Table \ref{one}. It is evident that the SM doesn't provide a good fit to the data as $\chi^2_{\rm min}\sim 28.14$ whereas for the $U_1$ LQ model, the fit is significantly improved as indicated by the $\chi^2_{\rm min}$ value which is $\sim $ 8.93 for $S1 \,\,\&\,\, S2$ and $\sim$ 11.55 for $S3$.
The 1$\sigma$ and 2$\sigma$ allowed regions of the new physics couplings are portrayed in Fig.~\ref{fig-ps}. 
The region in cyan color is the parameter space which is excluded by imposing the additional constraint of $B(B_c\rightarrow \tau \bar{\nu}) < 0.3$ \cite{Alonso:2016oyd}. One can see that the entire parameter space is excluded from $B(B_c\rightarrow \tau \bar{\nu})$ constraint for the $S3$ scenario.

%%%%%%%%%%%%%%%%%%%%%%%%%%%%%%%%%%%
\begin{figure*}[h]
\centering
\includegraphics[width = 2.1in]{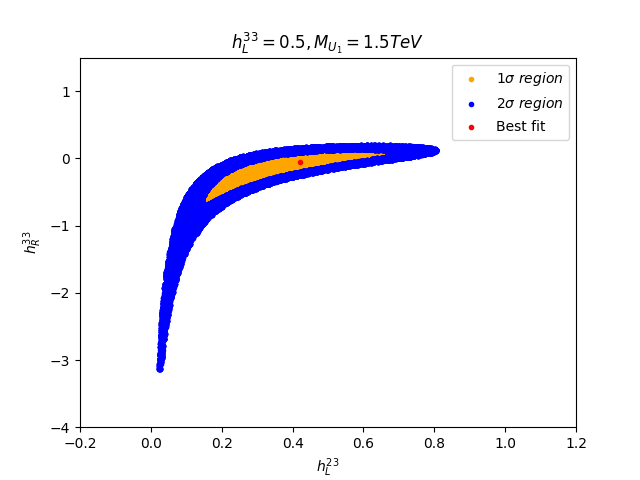}
\includegraphics[width = 2.1in]{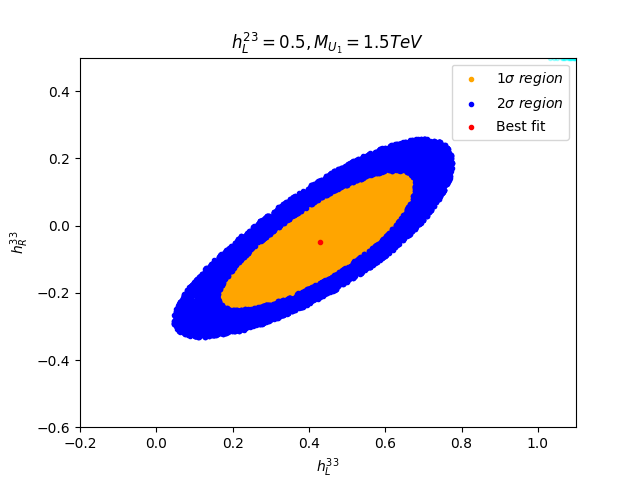}
\includegraphics[width = 2.1in]{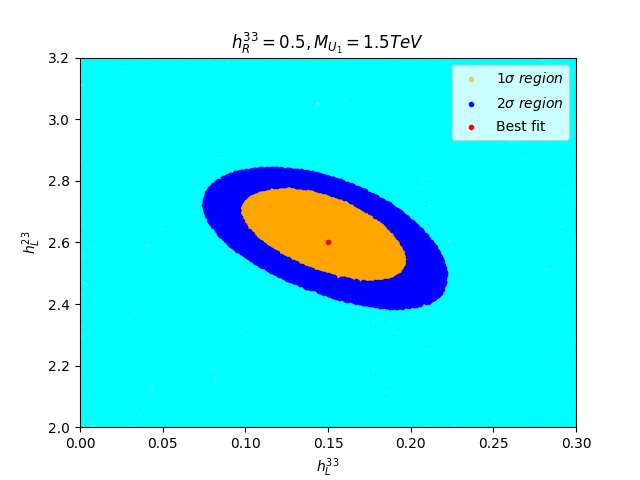}
\caption{The $1\sigma$ and $2\sigma$ allowed parameter space are shown in orange and blue color. The region with cyan color is the excluded region by imposing the constraint from $B(B_c\rightarrow \tau\bar{\nu})< 30 \%$.}
\label{fig-ps}
\end{figure*}

{\rowcolors{2}{black!50!white!50}{black!50!white!40}

\begin{table}
\centering
\begin{tabular}{|c|c|c|}
\hline\hline
 & Best fit value(s) & $\chi^2_{\rm min}$  \\
SM  & $C_{i}=0$ & 28.14 \\
$S1$  &  $h^L_{23} = 0.42 \pm 0.18, \, h^R_{33} =   -0.06    \pm    0.18$ & 8.93  \\
\hline
$S2$  & $h^L_{33} = 0.43 \pm 0.17, \, h^R_{33} =   -0.06   \pm    0.14$ & 8.93 \\
$S3$  & $h^L_{33} = 0.15 \pm 0.03, \, h^L_{23} =   2.64    \pm    0.10$ & 11.55\\
\hline  
\end{tabular}
\caption{ Best fit values of new physics couplings by making use of data of $R_D$, $R_{D^*}$, $R_{J/\Psi}$, $R_{\Lambda_c}$, $P_{\tau}^{D^*}$ and $f_L^{D^*}$ in the fit.}
\label{one}
\end{table}

Using the allowed values of the new physics couplings obtained in this section, in the next subsection, we predict several observables in the decay of $\Lambda_b \to p l\bar{\nu}$ for benchmark scenarios  NP(S1) with $h_L^{23} = 0.69, h_R^{33} = 0.09 $ and NP(S2) with $h_L^{33} = 0.67, h_R^{33} = 0.12$ which correspond to the maximum deviation from the SM predictions in the $1\sigma$ favoured new physics parameter space.

\subsection{Predictions}

\begin{figure*}[htb]
\centering
\includegraphics[width = 3.1in]{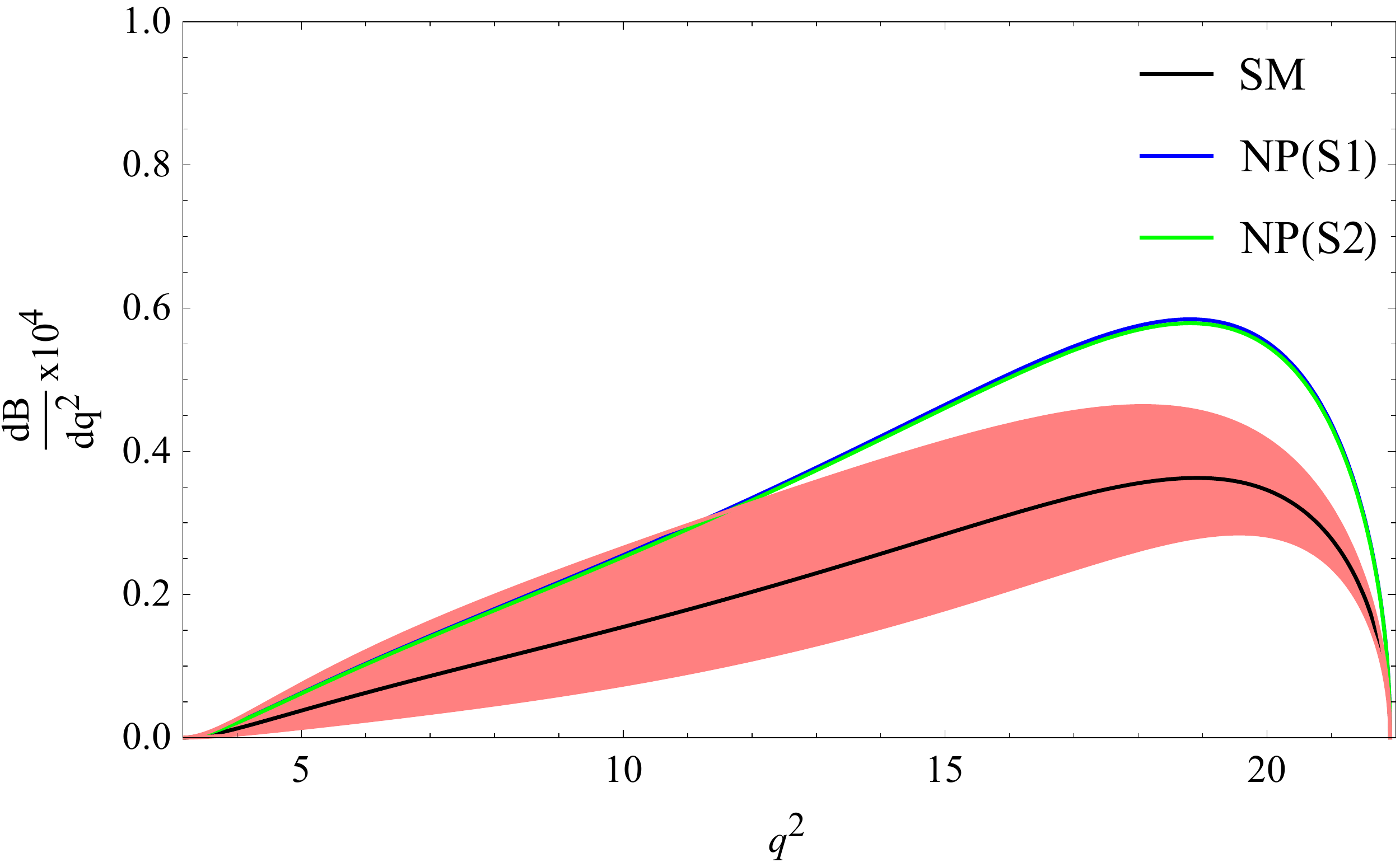}
\includegraphics[width = 3.1in]{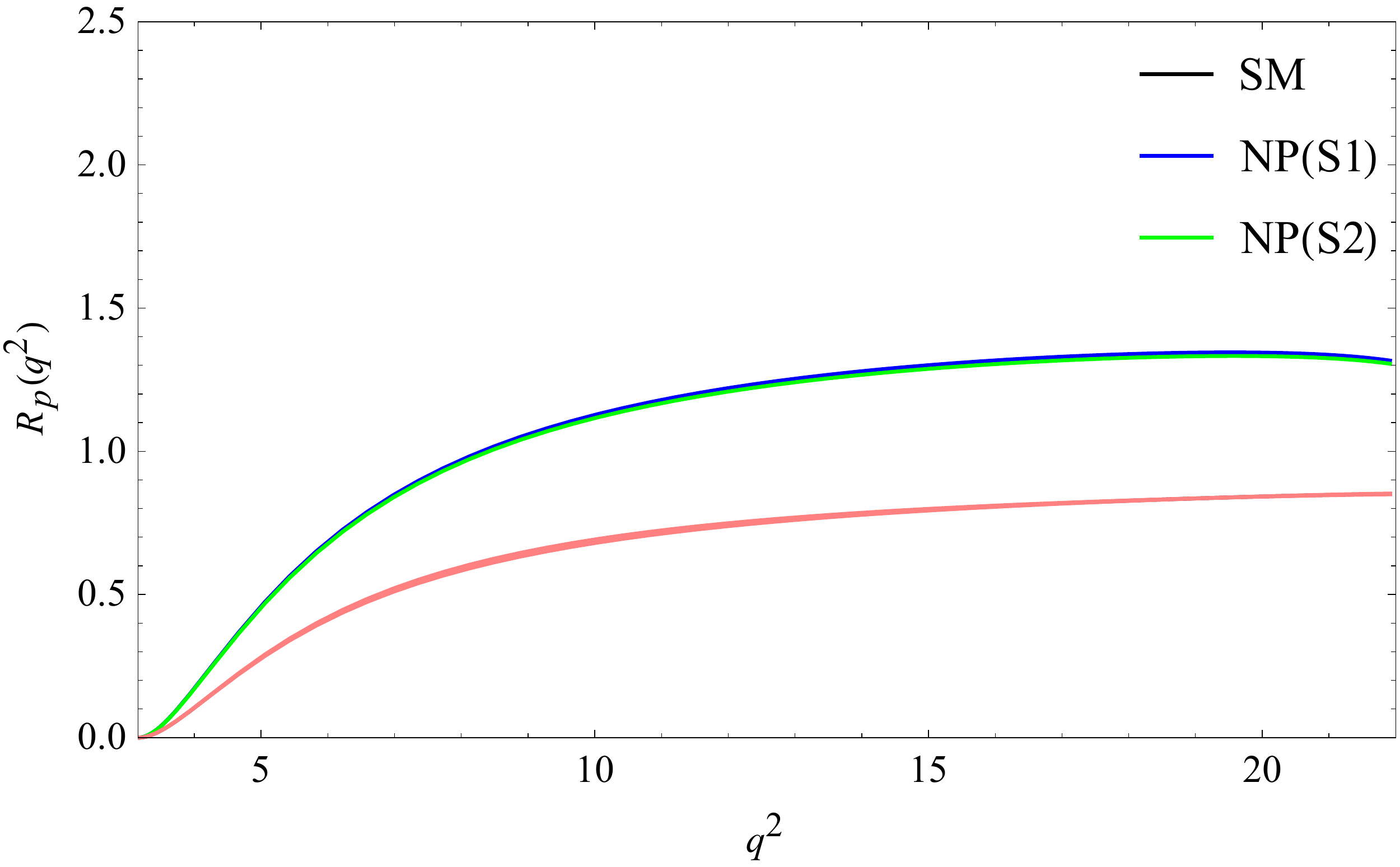}\\
\includegraphics[width = 3.1in]{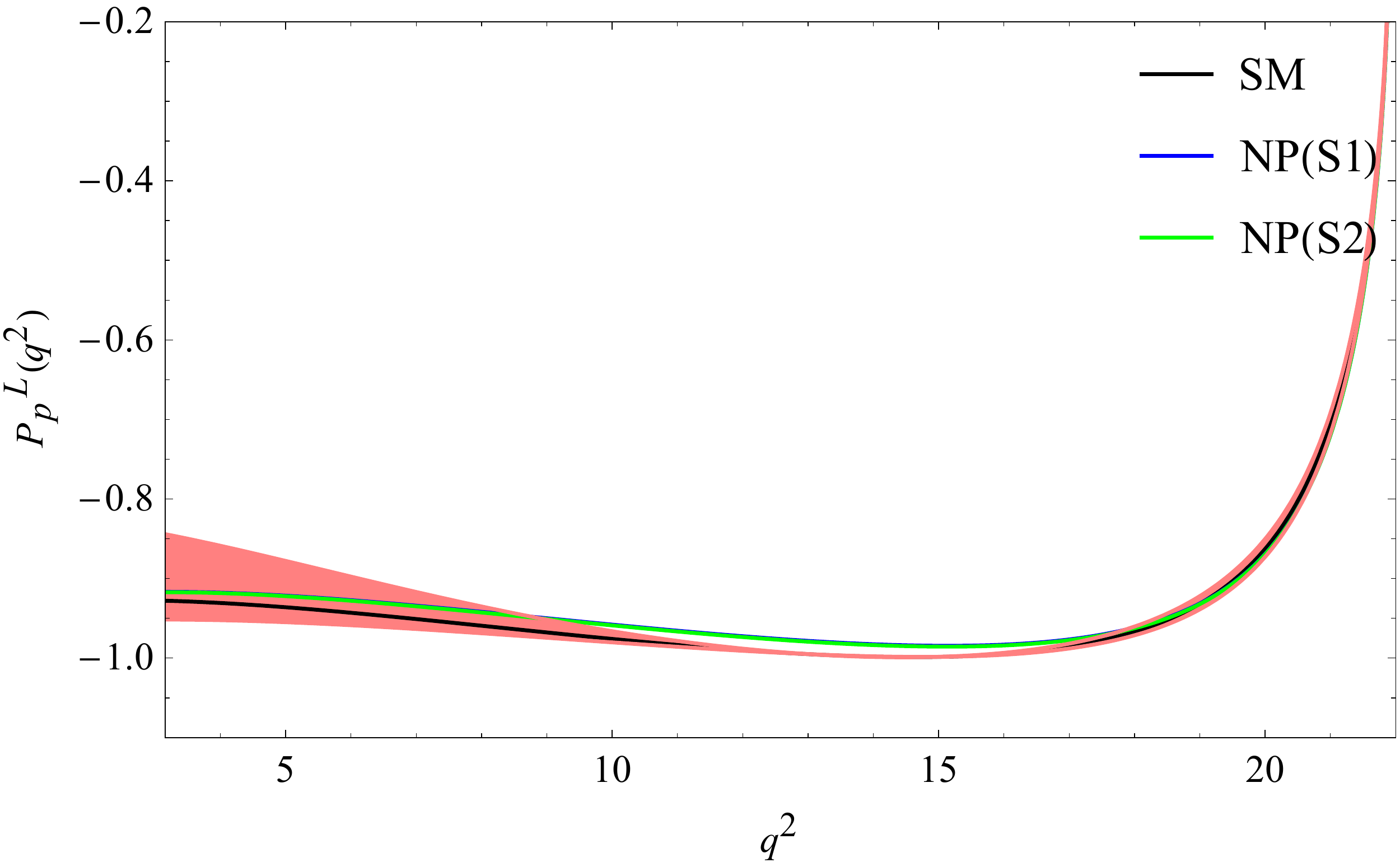}
\includegraphics[width = 3.1in]{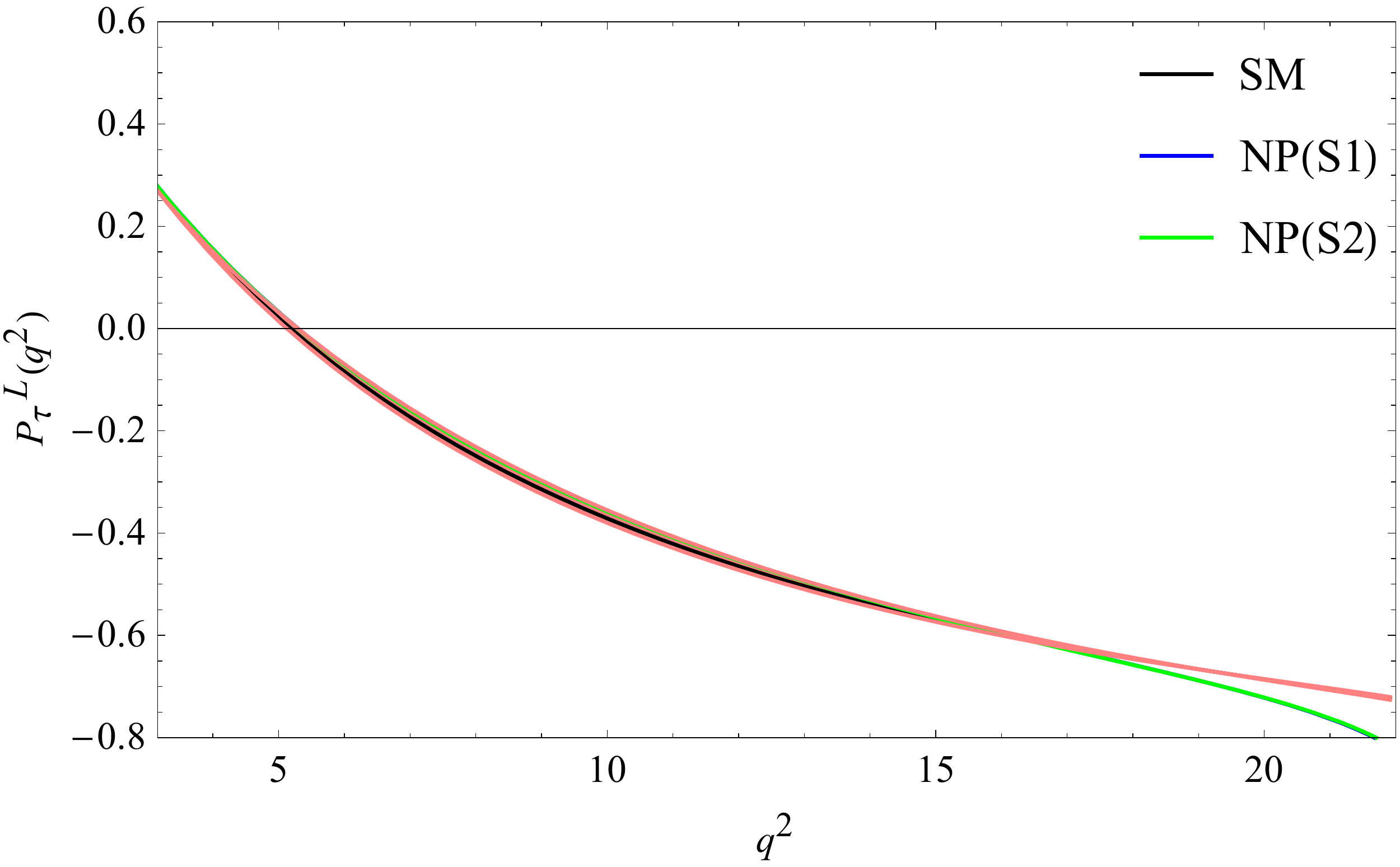}\\
\includegraphics[width = 3.1in]{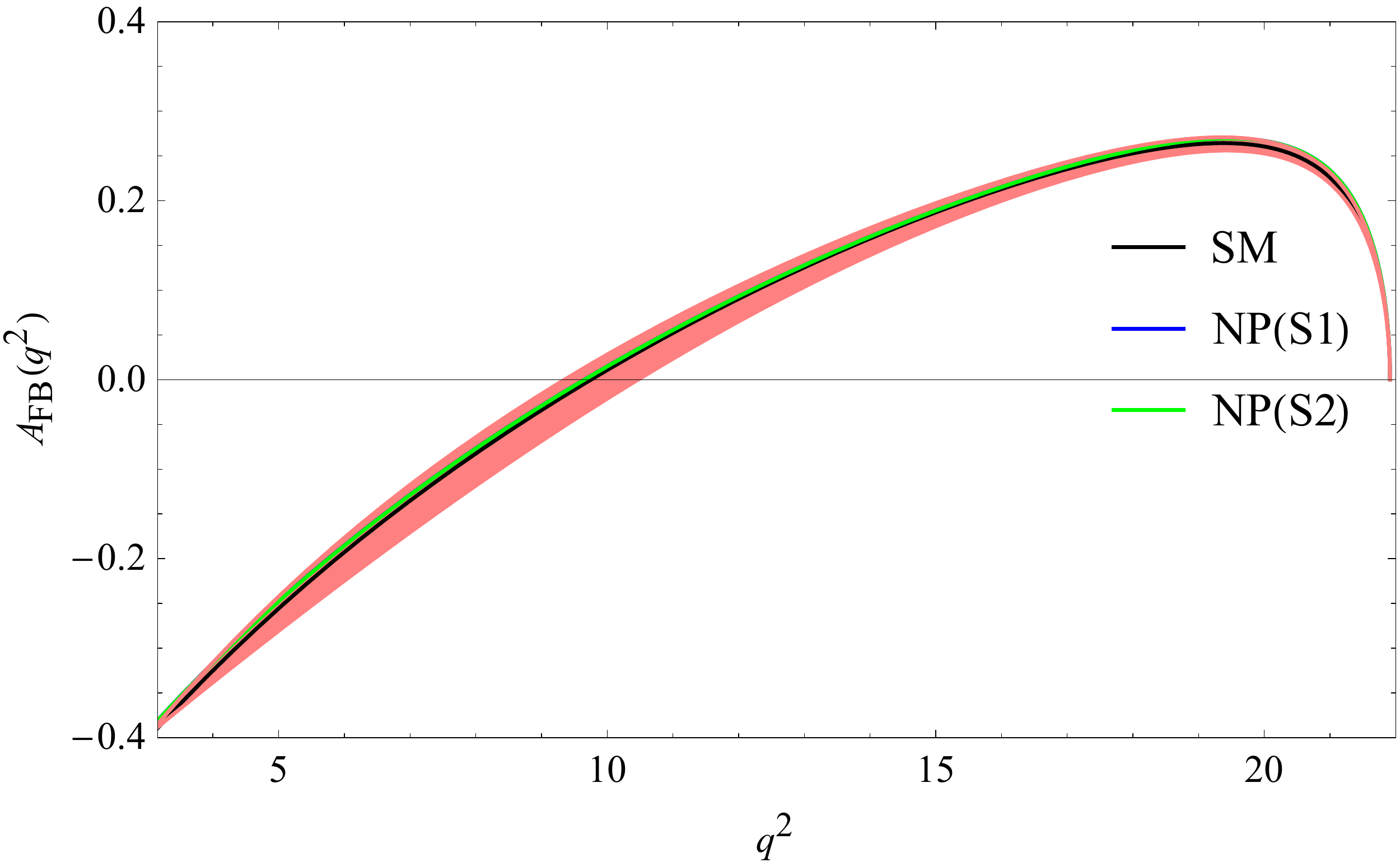}
\includegraphics[width = 3.1in]{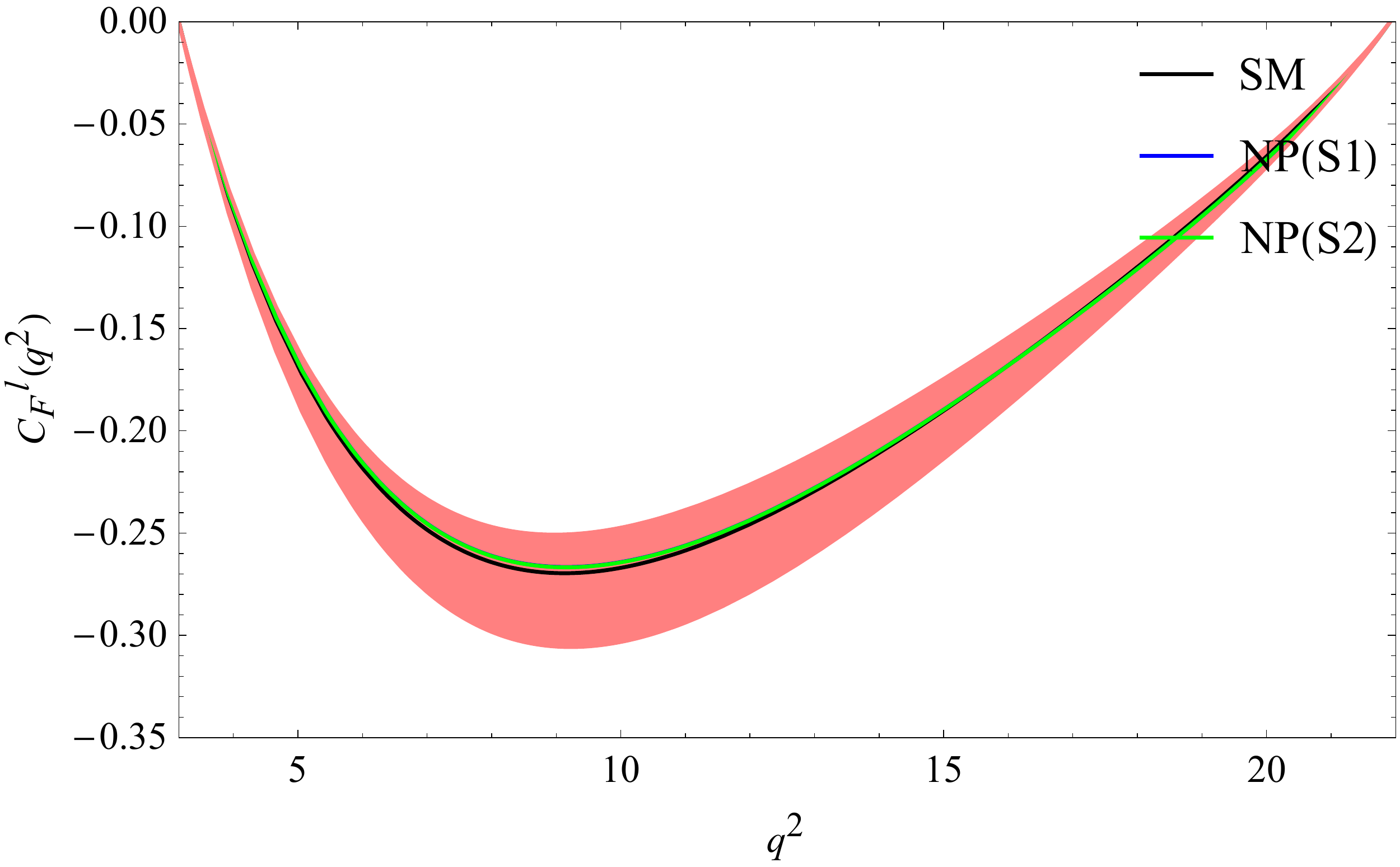}
\caption{Predictions for various observables in $\Lambda_b \to p l\bar{\nu}$ decay. The band  corresponds to the SM uncertainties. The lines in blue and green correspond to the maximum deviation from the SM predictions in the $1\sigma$ favoured new physics parameter space}.
\label{fig-pred}
\end{figure*}

We consider following $\Lambda_b \to p l\bar{\nu}$ observables in our analysis:
\begin{itemize}
\item differential branching ratio $dB/dq^2$, defined in eq. \eqref{dbr}

\item LFU ratio $R_p$, defined in defined in eq. \eqref{rlfu}

\item  longitudinal polarization of final state baryon,  defined in eq. \eqref{pol}

\item longitudinal polarization of $\tau$, defined in eq. \eqref{pol}

\item lepton forward-backward asymmetry $A_{FB}$, defined in eq. \eqref{afb}

\item convexity parameter $C_F^l$, defined in eq. \eqref{conv}. 
\end{itemize}

The SM prediction of these observables along with new physics benchmark scenarios NP(S1) and NP(S2) are illustrated in Fig.~\ref{fig-pred}. From the left panel of the top figure, it is luculent that the new physics can ameliorate the branching ratio by $\sim 2 $ times the SM prediction. Thus the current $b \to c$ data does allows an enhancement in the branching ratio of $\Lambda_b \to p l\bar{\nu}$. This roseate feature is also toted to the LFU ratio $R(\Lambda_p)$ as can be seen from the right panel of the top figure where about 2 times of magnitude enhancement is allowed. Therefore $\Lambda_b \to p l\bar{\nu}$ decay mode can serve as an important channel to probe LFU violation in the $b \to u $ sector.

 The predictions of the longitudinal polarization of the  final state baryon as well as the $\tau$ lepton in $\Lambda_b \to p l\bar{\nu}$ decay are shown in the left and right panels of Fig.~\ref{fig-pred}. For $q^2 <$ 8 $\rm GeV^2\,\,  \& >$  17 $\rm GeV^2$, $P^L_{p}(q^2) $ is consistent with the SM prediction whereas for 8 $\rm GeV^2 <$   $q^2<$ 17 $\rm GeV^2$, there is marginal deviation from the SM. On  the other hand, the predictions of tau polarization is consistent with the SM value in the entire $q^2$ region. The same is true for lepton forward backward asymmetry and convexity parameter as can be seen from the left and right panels of bottom figure, respectively.

\section{Conclusions}
\label{concl}

In this work we anatomize new physics effects in  $\Lambda_b \to p \tau \bar{\nu}$ decay in $U_1$ leptoquark model. This decay mode is induced by the quark level transition $b \to u \tau \bar{\nu}$. A model independent analysis of new physics in $b \to u \tau \bar{\nu}$ can lead to large effects due to the fact that, as of now, we only have one measurement in this sector. However, in the context of $U_1$ leptoquark model considered in this work, the new physics couplings in  $b \to u \tau \bar{\nu}$ transition can be expressed in terms of couplings in $b \to c \tau \bar{\nu}$ decay along with a suitable combinations of elements of the CKM matrix. Therefore, one expects a strong correlations between these two sectors. Given the fact that, unlike $b \to u \tau \bar{\nu}$ sector, there are measurements of a number of observables in decays induced by $b \to c \tau \bar{\nu}$ transition, one expects that  meaningful constraints on new physics parameter space can be obtained. It would then be interesting to see whether such constraints can allow for large enhancements in some of the observables in $\Lambda_b \to p \tau \bar{\nu}$ decay. 

In order to obtain constraints on new physics couplings, we perform a fit to all  $b \to c \tau \bar{\nu}$ data. For allowed parameter space of the couplings,  we obtain predictions of the branching ratio, LFU ratio, the longitudinal polarization of final state baryon and $\tau$ lepton, lepton forward-backward asymmetry and in the decay of $\Lambda_b \to p \tau \bar{\nu}$. We find that

\begin{itemize}
\item The branching ratio as well as the LFU ratio can be enhanced by about 2 times over the SM value.
\item There can be a marginal deviation from the SM in the longitudinal polarization of final state baryon for 8 $\rm GeV^2 <$   $q^2<$ 17 $\rm GeV^2$.
\item The longitudinal polarization of $\tau$, lepton forward-backward asymmetry as well as the convexity parameter are consistent with the SM. 
\end{itemize}

{\bf Acknowledgements}: The work of DK is supported by the SERB, India under the research grant no. SERB/EEQ/2021/000965. DK would like to thank Ashutosh Kumar Alok for useful discussion during this work.

\section{APPENDIX}
\subsection{$\Lambda_b \to p$ transition form factors and Helicity amplitudes}
\label{appen}

The $q^2$ dependence of the helicity form factors in the lattice QCD calculations are defined as\cite{Detmold:2015aaa}:
\begin{equation}
f_i(q^2) = \dfrac{1}{1-q^2/(m^f_{\small pole})^2}[a_0^f+a_1^f z(q^2)]\,,
\end{equation}
where $i = +, \perp, 0$ and the expansion parameter is defined as
\begin{equation}
z(q^2) = \dfrac{\sqrt{t_+-q^2}-\sqrt{t_+-t_0}}{\sqrt{t_+-q^2}+\sqrt{t_+-t_0}}\,.
\end{equation}
Here $t_+ = (m_{B_1} + m_{B_2})^2$ and $t_0 = (m_{B_1} - m_{B_2})^2$. The nominal form factor parameters $a_{0,1}^{f(g)}$ and $m^f_{pole}$ for $\Lambda_b \to p$ are taken from \cite{Detmold:2015aaa}.

The decay $\Lambda_b \to p l\bar{\nu}$ is considered to be through $\Lambda_b \to pW^*$ and the off-shell $W^*$ decays to $l\bar{\nu}$. The helicity amplitudes for vector and axial-vector type current is defined by

\begin{eqnarray}
H_{\lambda_p, \lambda_W} &=& H^V_{\lambda_p,\lambda_W} - H^A_{\lambda_p,\lambda_W}\,,\\
H^V_{\lambda_p,\lambda_W} &=& \epsilon^{\dagger\mu}(\lambda_W)\braket{p,\lambda_p|\bar{c}\gamma_{\mu}b|\Lambda_b, \lambda_{\Lambda_b}}\,,\\
H^A_{\lambda_p,\lambda_W} &=& \epsilon^{\dagger\mu}(\lambda_W)\braket{p,\lambda_p|\bar{c}\gamma_{\mu}\gamma_5 b|\Lambda_b, \lambda_{\Lambda_b}}\,.
\end{eqnarray}

Also, the helicity amplitudes for scalar and pseudo-scalar current is given by
\begin{eqnarray}
H^S_{\lambda_p} &=& \braket{p,\lambda_p|\bar{c}b|\Lambda_b, \lambda_{\Lambda_b}}\,,\\
H^P_{\lambda_p} &=& \braket{p,\lambda_p|\bar{c}\gamma_5 b|\Lambda_b, \lambda_{\Lambda_b}}\,.
\end{eqnarray}
One can show from the parity argument or explicit calculation that $H^V_{-\lambda_p, -\lambda_W} = H^V_{\lambda_p, \lambda_W}$, $H^A_{-\lambda_p, -\lambda_W} = -H^A_{\lambda_p, \lambda_W}$, $H^S_{\lambda_p, \lambda_{NP}} = H^S_{-\lambda_p, -\lambda_{NP}}$ and $H^P_{\lambda_p, \lambda_{NP}} = -H^P_{-\lambda_p, -\lambda_{NP}}$.

The helicity amplitudes can be defined in terms of the helicity form factors as\cite{Shivashankara:2015cta}:
\begin{eqnarray}
H^V_{\small{\frac{1}{2},0}} &=& (1+C_{V_L}+C_{V_R}) \dfrac{\sqrt{Q_-}}{\sqrt{q^2}}(m_{B_1}+m_{B_2})f_{+}(q^2)\,,\\
H^A_{\small{\frac{1}{2}},0} &=& (1+C_{V_L}-C_{V_R})\dfrac{\sqrt{Q_+}}{\sqrt{q^2}}(m_{B_1}-m_{B_2})g_+(q^2)\,,\\
H^V_{\small{\frac{1}{2}},1} &=& - (1+C_{V_L}+C_{V_R}) \sqrt{2Q_-}f_\perp(q^2)\,,\\
H^A_{\small{\frac{1}{2}},1} &=& - (1+C_{V_L}-C_{V_R}) \sqrt{2Q_+}g_\perp(q^2)\,,\\
H^V_{\small{\frac{1}{2}},t} &=& (1+C_{V_L}+C_{V_R}) \dfrac{\sqrt{Q_+}}{\sqrt{q^2}}(m_{B_1}-m_{B_2})f_0(q^2)\,,\\
H^A_{\small{\frac{1}{2}},t} &=& (1+C_{V_L}-C_{V_R}) \dfrac{\sqrt{Q_-}}{\sqrt{q^2}}(m_{B_1}+m_{B_2})g_0(q^2)\,,
\end{eqnarray}
where
\begin{equation}
Q_{\pm} = (m_{B_1} \pm m_{B_2})^2-q^2\,.
\end{equation}

The scalar and pseudo-scalar helicity amplitudes are defined as:
\begin{eqnarray}
H^{SP}_{\small{\frac{1}{2}},0} &=& H^S_{\small{\frac{1}{2}}0}-H^P_{\small{\frac{1}{2}}0}\,,\\
H^S_{\small{\frac{1}{2}},0} &=& (C_{S_L}+C_{S_R}) \dfrac{\sqrt{Q_+}}{m_b-m_{u}}(m_{B_1}-m_{B_2})f_0(q^2)\,,\\
H^P_{\small{\frac{1}{2}},0} &=& (C_{S_L}-C_{S_R}) \dfrac{\sqrt{Q_-}}{m_b+m_{u}}(m_{B_1}+m_{B_2})g_0(q^2)\,.
\end{eqnarray}

The helicity-dependent differential decay rates are required to compute the longitudinal polarization asymmetry of final state baryon and $\tau$ and these decay rates are defined as
\begin{eqnarray}
\frac{d\Gamma^{\lambda_p = \frac{1}{2}}}{dq^2} &=& \frac{m_l^2}{q^2}\Big[\frac{4}{3}\Big(H^2_{\small{\frac{1}{2},1}} + H^2_{\small{\frac{1}{2}0}} + 3H^2_{\small{\frac{1}{2},t}}\Big)\Big] + \frac{8}{3}\Big(  H^2_{\small{\frac{1}{2},0}} +  H^2_{\small{\frac{1}{2},1}}\Big) + 4 H^{SP^2}_{\small{\frac{1}{2},0}} +  \frac{8m_l}{\sqrt{q^2}}H_{\small{\frac{1}{2},t}} H^{SP}_{\small{\frac{1}{2},0}}\\
\frac{d\Gamma^{\lambda_p = -\frac{1}{2}}}{dq^2} &=& \frac{m_l^2}{q^2}\Big[\frac{4}{3}\Big(H^2_{\small{-\frac{1}{2},1}} + H^2_{\small{-\frac{1}{2},0}} + 3H^2_{\small{-\frac{1}{2},t}}\Big)\Big] + \frac{8}{3}\Big(  H^2_{\small{-\frac{1}{2},0}} +  H^2_{\small{-\frac{1}{2},-1}}\Big) + 4 H^{SP^2}_{\small{-\frac{1}{2},0}} \nonumber\\
&&+  \frac{8m_l}{\sqrt{q^2}}H_{\small{-\frac{1}{2},t}} H^{SP}_{\small{-\frac{1}{2},0}}\\
\frac{d\Gamma^{\lambda_{\tau} = \frac{1}{2}}}{dq^2} &=& \frac{m_l^2}{q^2}\Big[\frac{4}{3}\Big(H^2_{\small{\frac{1}{2},1}} + H^2_{\small{\frac{1}{2},0}} + H^2_{\small{-\frac{1}{2},-1}}  + H^2_{\small{-\frac{1}{2},0}}\Big) + 4(  H^2_{\small{\frac{1}{2},t}} +  H^2_{\small{-\frac{1}{2},t}}\Big)\Big]+ 4\Big( H^{SP^2}_{\small{\frac{1}{2},0}} + H^{SP^2}_{\small{-\frac{1}{2},0}}\Big)\nonumber\\
&&+  \frac{8m_l}{\sqrt{q^2}}\Big(H_{\small{\frac{1}{2},t}} H^{SP}_{\small{\frac{1}{2},0}} + H_{\small{-\frac{1}{2},t}} H^{SP}_{\small{-\frac{1}{2},0}}\Big)\\
\frac{d\Gamma^{\lambda_{\tau} = -\frac{1}{2}}}{dq^2} &=& \frac{8}{3}\Big(H^2_{\small{\frac{1}{2},1}} + H^2_{\small{\frac{1}{2},0}} + H^2_{\small{-\frac{1}{2},-1}}  + H^2_{\small{-\frac{1}{2},0}}\Big)
\end{eqnarray}

%%%%%%%%%%%%%%%%%%%%%%%%%%%%%%%%%%%%%%%%

\end{document}